\documentclass[twocolumn,amsmath,aps,superscriptaddress,showpacs,floatfix,a4paper]{revtex4}
\usepackage{amsmath}
\usepackage{graphicx}
\usepackage{epsf}
\usepackage{bm}
\usepackage{latexsym}
\RequirePackage{amsopn}
\RequirePackage{amssymb}
\RequirePackage{amsfonts}
\RequirePackage{amsthm}
\usepackage[]{graphicx}
\usepackage[]{amsmath}

\begin{document}

\title{No free lunch for effective potentials: general comment for Faraday FD144 }
\author{Ard A.\ Louis}
\affiliation{Rudolf Peierls Centre for Theoretical Physics,
           1 Keble Road, Oxford OX1 3NP, United Kingdom}
\begin{abstract}    I briefly review the problem of representability --- a single coarse-grained effective pair potential cannot simultaneously represent all the properties of an underlying more complex system -- as well as a few other subtleties that can arise in interpreting coarse-grained potentials.
\end{abstract}
\date{\today}

\maketitle

\section{Introduction} 

An important source of progress in computer simulation comes
from better coarse-grained models of the underlying materials, that is
descriptions that are simpler and more tractable, but nevertheless
retain the fundamental underlying physics that one is interested in
investigating~\cite{Prap08,Noid08,Murt09}.    To most of us it is intuitively obvious that these simplified descriptions throw some information away and that compromises are made. After all, there is no such thing as a free lunch.  But how much are we paying, and what can we get away with?

\subsection{coarse-graining: structure and  energy routes}

To illustrate these questions, consider the example of   
a one-component reference fluid interacting with a three-body Hamiltonian of the form:
\begin{equation}\label{eq2.1}
H = K + \sum_{i < j}w^{(2)}(r_{ij})
 + \sum_{i <j <k}w^{(3)}({\bf r}_{ij},{\bf r}_{jk},{\bf r}_{ki}),
\end{equation}
where ${\bf r_i}$ denotes the position of particle $i$ and ${\bf r}_{ij} = {\bf r}_i-{\bf r}_j$ and $r_{ij} = |{\bf
r}_i-{\bf r}_j|$.  $K$ is the kinetic energy operator, $w^{(2)}(r)$ is an
isotropic pairwise additive potential, and $w^{(3)}({\bf r}_{ij},{\bf
r}_{jk},{\bf r}_{ki})$ is a triplet or three-body potential.  Three-body potentials are expensive and
cumbersome to simulate, and so one might want to coarse-grain them to
a simpler isotropic representation.

There are several ways you could do this.   One popular method is to fit a pair-potential such that it reproduces a structural quantity like the  radial distribution function $g(r)$ generated by the Hamiltonian of Eq.~\ref{eq2.1}. 
Henderson~\cite{Hend74} first showed  that ``the pair
potential $v(r)$ which gives rise to a radial distribution function
$g(r)$ is unique up to a constant.''. A more rigourous mathematical discussion
is provided by Chayes, Chayes and Lieb~\cite{Chay84}.
 An extended proof for
orientational correlations can be found in a book by Gray and
Gubbins~\cite{Gray84} and we did  this for multi-site potentials in ref.~\cite{John07}.   Therefore, for
a given state-point, the $g(r)$ generated by a Hamiltonian like that of Eq.~(\ref{eq2.1})
can be reproduced by a {\em unique} effective pair potential $v_g(r)$.  (The existence of $v_g$ is more subtle, but holds under fairly general conditions, see ~\cite{Chay84}).  
I'll call approaches that derive $v_g(r)$ the {\em structural route} to deriving an effective potential.
The difference with the bare-pair potential $w^{(2)}(r)$ can be written as:
\begin{equation}\label{eq2.2b}
\delta v_g(r) = w^{(2)}(r) - v_g(r).
\end{equation}

Another popular way of  deriving an effective potential is by fitting to a thermodynamic observable like the 
internal energy.  One first calculates the full internal energy  $U(N,V,T)$ for the system governed by the Hamiltonian~(\ref{eq2.1}), and then (in this case) derives a potential $v_U^{eff}(r)$ that reproduces the same internal energy by the simpler two-body formula:
\begin{equation}\label{eq2.3b}
 U(N,V,T) = \frac12 \rho^2 \int d{\bf r}_1 d{\bf r}_2
 g(r_{12})
 v_{U}^{\rm eff}(r_{12}).
\end{equation}
I'll call this  the {\em energy route}.
The difference with the bare-pair potential $w^{(2)}(r)$ can be written as:
\begin{equation}\label{eq2.2c}
\delta v_U(r) = w^{(2)}(r) - v_U(r).
\end{equation}

It is not hard to show that both $v_g(r)$ and $v_U(r)$ depend on the
state point at which they are derived.  Thus if they are used at a
different state-point, one would expect {\bf transferability problems}, i.e. you'd need to re-derive them for your new state-point.

\subsection{representability problems}

What is perhaps more worrying is that one can also show that at a
given state-point, $v_g(r)$ and $v_U(r)$ cannot be the same.
As first demonstrated almost 40 years ago~\cite{Casa70}, to lowest order in $\rho$ and $w^{(3)}$, the ratio between the
two corrections is:
\begin{equation}\label{eq2.9}
\frac{\delta v_U(r)}{\delta v_g(r)} = \frac13 + {\cal
 O}\left((w^{(3)})^2;\rho^2\right).
\end{equation}
In other words  even in the low density weak potential limit, the corrections due to the three-body forces that you are coarse-graining out differ by a factor of three!  What happens for stronger interactions or higher densities needs to be investigated.
Since $v_g(r)$ is unique, it is therefore impossible to represent all the 
properties of a system governed by the Hamiltonian of
Eq.~(\ref{eq2.1}) by a single pair potential.  There is no free lunch.  I believe such  {\bf
representability problems} are widespread in coarse-grained
descriptions of soft-matter systems\cite{Loui02}.

They may also be important
important in other coarse-grained simulations. For example, we
recently used the structural route to derive radially symmetric
potentials for water from a more sophisticated underlying model~\cite{John07}.  We explicitly constructed  $v_g(r)$ and $v_U(r)$ and they look very different, as anticipated by 
Eq.~{\ref{eq2.9}.  At some  of the state-points we studied, using $v_g(r)$ to
calculate the virial pressure resulted in a dimensionless compressibility factor $Z = \beta P/\rho$
that was almost two orders of magnitude larger than that of the
original multi-site water model used to parameterise $v_g(r)$. Similar deviations from the underlying water model were also seen in ref.~\cite{Wang09}.

Admittedly, it may not be surprising that an isotropic potential
should perform so poorly when the underlying fluid has complex orientational correlations.  In fact treating water as a pair potential is probably not such a good idea  

 One way of improving on the thermodynamic performance of the structurally derived potential $v_g(r)$ is to use constraints to simultaneously fit to properties like the virial pressure or the internal energy. 
 Due to the uniqueness of
$v_g(r)$, this potential would no longer correctly reproduce the pair
correlations, but since the thermodynamic properties are  scalar quantities, it might not come at too large a cost for the structure.  A nice example of applying this to water can be found in this paper by the Mainz group~\cite{Wang09} who showed that the structural properties were affected (and by extension the isothermal compressibility which can be written as an integral over $g(r)$), but not too badly.

\subsection{more general problems with effective potentials}

In ref.~\cite{Loui02} some more general issues with effective potentials are reviewed.
For example, most of the time our effective simplified potentials are really state dependent, that is if you were to derive them at different state points they would be different.     I argue that you shouldn't 
naively treat  them as if they are real two-body potentials of a Hamiltonian system.

Here is an example of how things can go horribly wrong. Consider a homogeneous fluid in a
volume $V$,  with $N$ particles interacting with a spherically symmetric
pair potential $v(r;\rho)$, that depends on the state point (that could be density, temperature, etc...). For simplicity, here we consider only the density
$\rho =N/V$. 
The standard way to
derive the virial equation  is directly through the {\em canonical}
partition function
\begin{equation}\label{eq2.3}  
Q(N,V,T) = \frac{\Lambda^{-3N}}{N!} \int d{\bf r}^N \exp \left\{ -\beta
\sum_{i<j} v(r_{ij};\rho) \right\}.
\end{equation}
where $\Lambda$ is the usual thermal de Broglie wavelength.
The volume derivative in 
\begin{equation}\label{eq2.4} 
\beta P = \left(\frac{\partial\log Q(N,V,T)}{\partial V} \right)_{N,T}
\end{equation}
should also act on $v(r;\rho)$, resulting in a virial equation 
with an extra $\partial v(r;\rho)/\partial \rho$  term:
\begin{eqnarray}\label{eq2.2}
Z_{vir}^\rho & = & \frac{\beta P}{\rho} \\ \nonumber 
 = & 1 - &  \frac{2}{3} \beta \pi \rho
\int_0^\infty r^2 \left\{r \frac{\partial v(r;\rho)}{\partial r} - 3
\rho \frac{\partial v(r;\rho)}{\partial \rho} \right\} g(r) dr,
\end{eqnarray}
something first pointed out in 1969 by Ascarelli and Harrison\cite{Asca69} in the context of density
dependent pair potentials used for modelling liquid metals.  

Now on the surface this all looks very kosher  -- I took my potential, plugged it into my partition function,  turned the statistical mechanical crank, and out popped Eq.~\ref{eq2.2} with a correction to the pressure due to the state-dependence of the effective potential.     However, when we tried this with the $v_g(r)$ derived for the water model~\cite{John07}, the correction actually made the agreement far worse.   In ref.~\cite{Loui02} I give other examples were this correction has the wrong sign (and magnitude).  
Although the analysis leading to Eq.~\ref{eq2.2} has the veneer of statistical mechanical respectability, it is in fact a disreputable result.

Taking a step back, it is not at all surprising that this derivation cannot be right~\cite{Loui02}.  The correction term in 
Eq.~\ref{eq2.2} only takes into account a local dependence on density.  If you took the same potential $v(r;\rho)$, and used it in a grand-canonical ensemble, then you would need information from all densities $\rho$, not just the local derivative.   So the two ensembles would not be equivalent, a good sign that there are more problems afoot.      Warnings about such difficulties abound, for example, about 40 years ago Barker {\em et al.}~\cite{Bark69} wrote:
\begin{quotation}{\em
We record our opinion that the use of density-dependent effective pair
potentials can be misleading unless it is recognised that these are
mathematical constructs to be used in specified equations rather than
physical quantities}
\end{quotation}
and similar reservations were sounded by other distinguished investigators~\cite{Casa70,Rowl84,Hoef99}.

The point of this exercise is not to say that effective potentials are useless. I've happily used them myself~\cite{Loui00}.  Sometimes  a fine solution to a dubious Hamiltonian is better than  a dubious solution to a very fine Hamiltonian.  It is just that one must be careful when using coarse-graining procedures not to automatically treat the coarse-grained potentials as if they are real bona-fide potentials of the type one uses in a Hamiltonian.    Also,    in practice it is also good to remember that these potentials are always compromises and that fitting too strongly to one quantity may generate larger errors in other quantities that one also wants to measure~\cite{Loui02,John07}.

\section{Questions for all}

\begin{enumerate}
\item  Do others have good examples of representability problems generated by their potentials?   I'd be interested to know how important they are in other systems.
\item A related question is:  How well can you predict the magnitude of a representability problem?   In the case of water it is not surprising that a radially symmetric pair-potential would have problems, but could you have predicted that a potential that generates $g(r)$ so well does so poorly on the internal energy or the virial pressure?
\item Ignacio Pagonabarraga has an interesting interpretation of density dependent potentials in ref.~\cite{Mera07}. Do you think this approach could be generalised to the kind of coarse-graining procedures we are applying in this conference?

\end{enumerate}

}

\end{document}